\documentclass{article}
\pdfoutput=1

\usepackage{amssymb,amsfonts,amsmath}
\usepackage{cite,enumerate,float}
\usepackage{color}
\usepackage{tikz}
\usetikzlibrary{arrows,snakes,backgrounds}
\usepackage[title]{appendix}

\def\be{\begin{eqnarray}}
\def\ee{\end{eqnarray}}

\def\p{\partial}

\def\Tr{{\rm Tr}\,}

\definecolor{red}{rgb}{1,0,0}
\definecolor{orange}{rgb}{1,0.5,0}
\definecolor{violet}{rgb}{0.7,0,1}

\usepackage{amssymb,amsfonts,amsmath,stmaryrd}
\usepackage{cite,enumerate,float,indentfirst}
\usepackage{color}
\usepackage{tikz}
\usetikzlibrary{arrows,snakes,backgrounds,calc}
\usepackage{ytableau}
\usepackage[vcentermath]{youngtab}

\newcommand\shadetext[2][]{%
  \setbox0=\hbox{{#2}}%
  \tikz[baseline=0]\path [#1] \pgfextra{\rlap{\copy0}} (0,-\dp0) rectangle (\wd0,\ht0);%
}
\newcommand{\rbtext}[1]{\shadetext[left color=blue, right color=red]{\bfseries #1 } }

\def\be{\begin{eqnarray}}
\def\ee{\end{eqnarray}}

\hoffset= - 1.0in         

\begin{document}

\title{\vspace{1.5cm}\bf
Interpolating Matrix Models for WLZZ series}

\author{
A. Mironov$^{b,c,d,}$\footnote{mironov@lpi.ru,mironov@itep.ru},
V. Mishnyakov$^{a,b,}$\footnote{mishnyakovvv@gmail.com},
A. Morozov$^{a,c,d,}$\footnote{morozov@itep.ru},\\
\\
A. Popolitov$^{a,c,d,}$\footnote{popolit@gmail.com},
Rui Wang$^{e,}$\footnote{wangrui@cumtb.edu.cn},
Wei-Zhong Zhao$^{f,}$\footnote{zhaowz@cnu.edu.cn}
}

\date{ }

\maketitle

\vspace{-6.5cm}

\begin{center}
\hfill FIAN/TD-01/23\\
\hfill IITP/TH-01/23\\
\hfill ITEP/TH-01/23\\
\hfill MIPT/TH-01/23
\end{center}

\vspace{4.5cm}

\begin{center}
$^a$ {\small {\it MIPT, Dolgoprudny, 141701, Russia}}\\
$^b$ {\small {\it Lebedev Physics Institute, Moscow 119991, Russia}}\\
$^c$ {\small {\it ITEP, Moscow 117218, Russia}}\\
$^d$ {\small {\it Institute for Information Transmission Problems, Moscow 127994, Russia}}\\
$^e${\em Department of Mathematics, China University of Mining and Technology,
Beijing 100083, China}\\
$^f${\em School of Mathematical Sciences, Capital Normal University,
Beijing 100048, China} \\
\end{center}

\vspace{.0cm}

\begin{abstract}
We suggest a two-matrix model depending on three (infinite) sets of parameters
which interpolates between all the models proposed in \cite{WLZZ}
and defined there through $W$-representations.
We also discuss further generalizations of the WLZZ models,
realized by $W$-representations associated with infinite commutative families of generators of $w_\infty$-algebra
which are presumably related to more sophisticated multi-matrix models.
Integrable properties of these generalizations are described by
what we call the skew hypergeometric $\tau$-functions.
\end{abstract}

\section{Introduction}

Recently in \cite{WLZZ} a new class of $\tau$-functions was introduced with the help of $W$-representations
generated by an extended class of operators.
These $\tau$-functions possess superintegrability property \cite{MMsi},
but more standard features of the (eigenvalue) matrix model partition functions \cite{UFN3} are not immediately
obvious: those from the Virasoro-like constraints to topological ($1/N$) expansion \cite{MMsc}.
Moreover, at least half of these models (``positive branch") seem to violate the usual relation
between these features: the AMM/CEO topological recursion \cite{AMM,EO} originating
from the Virasoro/W-algebra constraints expanded around a given spectral curve \cite{AMMdop}
seem to be related to $N$ rather than to $1/N$ expansion \cite{MMsc}.

The goal of this  letter is to begin unraveling the intriguing mysteries of the WLZZ models.
Our first claim is that these models are indeed the matrix models:
we suggest explicit two-matrix model integrals for them, which reduce to single matrix integrals
in particular cases (like the $n=\pm 2$ members of the WLZZ model sequences).
The second claim is that the class of the models can be further extended to include other $W$-operators
from a Borel subalgebra of the $w_{\infty}$-algebra.
The third claim is that though all these models are naturally united into
just two branches, positive and negative ones, and, though the first one is a particular case of the second one,
the $W$-representations for these two branches still need to be treated differently.

This reformulation of the story looks quite illuminating and should be a nice starting point
for a serious study of this unifying approach to non-perturbative partition functions.

The structure of the letter is as follows. In sec.2, we introduce the most general partition function of the type we are interested in: this type of partition function is associated with a $\tau$-function of the Toda lattice hierarchy \cite{UT} and generalizes hypergeometric $\tau$-functions introduced earlier in \cite{GKM2,OS}.
We call such $\tau$-functions skew hypergeometric.
Surprisingly, even for this rather general class of partition functions
there is a $W$-representation, though it is realized by the $W$-operators
which are not always easy to find in an explicit form.

In sec.3, we specify further the partition functions to the class that contains the whole set of the WLZZ models. These partition functions can be realized by a two-matrix integral that depends on two sets of variables $\bar p_k$ and $g_k$ and on an external matrix $\Lambda$, and interpolates between all the WLZZ models. This matrix integral can be realized by a $W$-representation given by explicit differential operators in variables $p_k=\Tr\Lambda^k$. Moreover, the matrix integral that describes a particular case when all $p_k=\Lambda=0$ (which interpolates between the WLZZ models of the negative branch) can be also realized by another $W$-representation.

In sec.4, we discuss an extension of the WLZZ model series, and discuss its $W$-representation. It turns out that all $W$-representations discussed in the letter are associated with generators from a Borel subalgebra of the $w_{\infty}$-algebra.

Sec.5 contains some concluding remarks. In the Appendices, we also added technically important issues related to action of generators of the $W$-representations on the basis of the Schur functions, and some simple examples of these generators.

\paragraph{Notation.} We use the notation $S_R\{p\}$ for the Schur functions, which are graded polynomials of arbitrarily many variables $p_k$ of grading $k$. The Schur function $S_R\{p\}$ is labelled by the Young diagram (partition) $R$: $R_1\ge R_2\ge\ldots\ge R_{l_R}>0$, and the grading of this Schur function is $|R|:=\sum_iR_i$. Similarly, we denote $S_{R/P}\{p\}$ the skew Schur functions \cite{Mac}.

The Schur function $S_R\{p\}$ is a linear combination of monomials $p_\Delta:=\prod_i^{l_\Delta}p_i$ parameterized by the Young diagrams $\Delta$ such that $|\Delta|=|R|$.
One can also write the same monomial in the form $p_\Delta=\prod_{k=1} p_k^{m_k}$.
This latter parameterization of $p_\Delta$ is related to the quantity $z_\Delta:=\prod_k k^{m_k}m_k!$, which is the standard symmetric factor of the Young diagram (order of the automorphism).

We will use the scalar product $\left\langle\ldots \right\rangle$, which the standard Schur scalar product \cite{Mac} given by $\Big<p_\Delta\Big|p_{\Delta'}\Big>=z_\Delta\delta_{\Delta,\Delta'}$ and extended to the Schur functions by linearity. In particular,
\be\label{scp}
\Big<S_R\Big|S_Q\Big>=\delta_{R,Q},\ \ \ \ \ \Big<p_kS_R\Big|S_Q\Big>=\Big<S_R\Big|k{\p S_Q\over\p p_k}\Big>,\ \ \ \ \
\sum_\Delta {g_\Delta\over z_\Delta}\Big<p_\Delta\cdot S_Q\Big|S_R\Big>=S_{R/Q}\{g\}
\ee

\section{Skew hypergeometric $\tau$-functions and their integrable properties}

\subsection{Definition}

We start with considering the most general partition functions which we need in this paper. They are of the form
\be\label{Zf}
Z_f(\bar p,p,g)=\sum_{R,Q}{\prod_{i,j\in R} f(j-i)\over \prod_{i,j\in Q} f(j-i)}S_{R/Q}\{\bar p\}S_R\{g\}S_Q\{p\}:=
\sum_{R,Q}\prod_{i,j\in R/Q} f(j-i)S_{R/Q}\{\bar p\}S_R\{g\}S_Q\{p\}
\label{skewdef}
\ee
with some (arbitrary) function $f(x)$. Such a function $Z_f$ is not a hypergeometric $\tau$-function \cite{GKM2,OS,AMMNi}, unless $p_k=0$ and there are no skew functions in (\ref{skewdef}).
However, it turns out that it is still a $\tau$-function of the KP hierarchy w.r.t. the both sets of times\footnote{Note that the traditional choice $t_k$ of time variables of the KP hierarchy as compared with power sums $p_k$ of variables in symmetric functions is $t_k={p_k\over k}$.} $p_k$ and $g_k$. Hence, we call it {\bf skew hypergeometric $\tau$-function}.

Let us explain that, for an arbitrary function $f$, the partition function (\ref{Zf}) is a $\tau$-function w.r.t. to both $p_k$ and $g_k$ variables, moreover, making substitution $f(x)\to f(x+N)$, even a stronger statement is correct: $Z_f$ is a $\tau$-function of the Toda lattice hierarchy with $N$ being the Toda zeroth time. In order to prove this, let us note that, in accordance with \cite{Taka}, the sum
\be\label{Taka}
\tau_N(g,p|\xi)=\sum_{R,Q}\xi_{R,Q}(N)S_R\{g\}S_Q\{p\}
\ee
is a $\tau$-function of the Toda lattice hierarchy iff
\be\label{Takaxi}
\xi_{R,Q}(N)=\det_{i,j\le N}F(R_i-i,Q_j-j)
\ee
with some function $F$, and $N$ playing the role of the zeroth time.

Now we just remind that the skew Schur function has the Jacobi-Trudi determinant representation
\be
S_{R/Q}\{\bar p\}=\left\{\begin{array}{cl}
\det_{i,j\le l_R}h_{R_i-i-Q_j+j}\{\bar p\}&\hbox{if }Q\in R\cr
&\cr
0&\hbox{if }Q\notin R
\end{array}\right.
\ee
where $h_k$ are the complete homogeneous symmetric polynomials, that is, $h_k=S_{[k]}$. Thus, for the partition function (\ref{Zf}), we obtain representation (\ref{Taka}) with
\be
\xi_{R,Q}(N)=\left\{\begin{array}{cl}
\det_{i,j\le N}{G(R_i-i)\over G(Q_j-j)}h_{R_i-i-Q_j+j}\{\bar p\}&\hbox{if }Q\in R\cr
&\cr
0&\hbox{if }Q\notin R
\end{array}\right.
\ee
where
\be
G(k):=\prod_{i=1}^kf(N+i)
\ee

\subsection{$W$-representation}

Using $W$-representations for generating partition functions is the standard idea, which goes back to \cite{DJKM,wrep,wrep1,wrep2}, and, in application to matrix models, to \cite{MSh,Wmore,Max,MMM}. It turns out that the partition function (\ref{Zf}) also admits such a description.

As usual for matrix model partition functions, it can be described by the action
of exponential of a $W$-operator on exponential of time variables:
\be\label{Z+f}
Z_f(\bar p,p,g)=\exp\left(\sum_{k=1}{\bar p_m\hat W^f_m[p]\over m}\right)\cdot \exp\left(\sum_{k=1}{p_kg_k\over k}\right)
\ee
This operator acts on the variables $p_k$, and action of the commuting operators $\hat W^f_m[p]$ can be manifestly described in the basis of the Schur functions:
\be\label{Wm}
\hat W^f_m[p]S_R\{p\}= \sum_{R'} \prod_{i,j \in R/R'} f(j-i) \left\langle  \left.m{\p S_R\{p\}\over \p p_m}\right| S_{R'}\{p\} \right\rangle \,S _{R'}\{p\}
\ee

\subsection{Hypergeometric $\tau$-functions}

An interesting particular case of the partition function (\ref{Zf}) is at the point where all $p_k=0$. Then,  the partition function becomes
\be\label{Zfg}
Z_f(\bar p,p=0,g)=\sum_{R}\prod_{i,j\in R} f(j-i)S_{R}\{\bar p\}S_R\{g\}
\ee
It is a $\tau$-function of the hypergeometric type. However, in order to construct a $W$-representation, one can not just make a reduction of (\ref{Wm}), since this latter is an operator acting on variables $p_k$. Hence, in this case, one needs another $W$-representation, acting on variables $g_k$. Such a representation does exist, and is given by action of exponential of a $W$-operator on unity:
\be\label{Z+fg}
Z_f(\bar p,p=0,g)=\exp\left(\sum_{k=1}{\bar p_m\hat{W}^f_{-m}[g]\over m}\right)\cdot 1
\ee
This operator acts on the variables $g_k$, and action of the commuting operators $\hat{W}^f_{-m}[g]$ can be manifestly described in the basis of the Schur functions (meaning of negative subscript will become clear in the next section):
\be\label{W-m}
\hat{W}^f_{-m}[g] S_R\{g\} = \sum_{R'} \prod_{i,j \in R'/R} f(j-i) \left\langle g_m S_R\{g\}\Big| S_{R'}\{g\} \right\rangle \,S _{R'}\{g\}
\ee
With concrete choices of the function $f(x)$, as we shall demonstrate below, one can also construct these operators as very explicit differential operators.

\section{Interpolating matrix model for the WLZZ series}

\subsection{Interpolating partition function}

Let us specify to the case of a linear function $f(x)=N+x$:
\be\label{Z1}
Z(N;\bar p,p,g)=\sum_{R,Q}\prod_{i,j\in R/Q}(N+j-i)S_{R/Q}\{\bar p\}S_R\{g\}S_Q\{p\}
\ee
Choosing here $\bar p_k=\delta_{k,m}$, we arrive at the positive branch $Z_m$ of the WLZZ models \cite{WLZZ}:
\be
Z_m(N;p,g)=\sum_{R,Q}\prod_{i,j\in R/Q}(N+j-i)S_{R/Q}\{\delta_{k,m}\}S_R\{g\}S_Q\{p\}
\ee
Moreover, choosing further $p_k=0$, we arrive at the negative branch $Z_{-,m}$ of the WLZZ models \cite{WLZZ},
\be
Z_{-,m}(N;g)=\sum_{R}\prod_{i,j\in R}(N+j-i)S_{R}\{\delta_{k,m}\}S_R\{g\}
\ee

\subsection{Matrix model}

We propose that the model (\ref{Z1}), which interpolates between all the WLZZ models
and includes all of them at particular values of parameters, 
is described by the two-matrix model that depends on two (infinite) sets of variables $\bar p_k$, $g_k$ and an external matrix $\Lambda$:
\be\label{im}
\boxed{
Z(N;\bar p,p,g)=\int\int_{N\times N}dXdY\exp\left(\Tr XY-\Tr Y\Lambda+\sum_k {g_k\over k}\Tr X^k+\sum_k{\bar p_k\over k}\Tr Y^k\right)
}
\ee
with $p_k=\Tr\Lambda^k$.  Here the integration is understood as power series in $g_k$, $\bar p_k$ and $\Tr\Lambda^k$, and $X$ are Hermitian matrices, while $Y$ are anti-Hermitian ones. This means that this formula can be described in the pure combinatorics terms of Feynman diagrams, with the propagator being $\Big<X_{ij}Y_{kl}\Big>=\delta_{il}\delta_{jk}$.  One can make a thorough computer check that these Feynman diagrams, indeed, give the expansion (\ref{Z1}). A complete and direct derivation of (\ref{im}) will be provided elsewhere: it is more technical and can overshadow the simple pattern described in this letter. 

Let us check that particular cases are also correctly reproduced from this expression.

First of all, at $p_k=0$, i.e. at $\Lambda=0$, one obtains the interpolating model for the negative branch of the WLZZ models. This model is $Z_{2,1}$ in the notation of \cite{AMMN} and has a two-matrix model representation \cite{Orlov},\cite[Eq.(53)]{AMMN},\cite{Al}:
\be\label{nb}
Z_{-}(N;\bar p,g)=\int\int_{N\times N}dXdY\exp\left(\Tr XY+\sum_k {g_k\over k}\Tr X^k+\sum_k{\bar p_k\over k}\Tr Y^k\right)
\ee
which, indeed, follows from (\ref{im}) at $\Lambda=0$.

The second particular case emerges at $\bar p_k=\delta_{k,2}$, when one arrives at the particular WLZZ model of the positive branch at $m=2$. Indeed, in this case, (\ref{im}) reduces to formulas \cite[Eq.(2.40)]{WLZZ1}, \cite[Eq.(30)]{WLZZ}, \cite[Eqs.(93)-(94)]{MMsc} after performing the Gaussian integration over $Y$:
\be
Z_2=\int dX \exp\left(-{1\over 2}\Tr X^2+\sum_k {g_k\over k}\cdot\Tr (X+\Lambda)^k\right)
\ee

The third particular case is just the trivial one: at all $\bar p_k=0$. Then, integration over $Y$ gives the $\delta$-function, $\delta(X-L)$, and
\be
Z(N;0,p,g)=\exp\left(\sum_k{g_kp_k\over k}\right)
\ee
which, indeed, follows from (\ref{Zf}).

\subsection{$W$-representation}

With the concrete choice of the function $f(x)=x+N$ as in (\ref{Z1}), one can realize the $W$-representations operators (\ref{Wm}) and (\ref{W-m}) as differential operators. Operators $\hat W_m[p]$ from (\ref{Wm}) are constructed using three auxiliary operators: the cut-and-join operator \cite{GJ,MMN}
\be\label{W0}
\hat W_0=\hat W_0(N) =\dfrac{1}{2} \sum_{a, b}\left((a+b) p_a p_b \frac{\partial}{\partial p_{a+b}}+a b p_{a+b} \frac{\partial^2}{\partial p_a \partial p_b}\right) + N\sum kp_k \dfrac{\partial}{\partial p_k}
\ee
and operators
\be
\hat F_1=\Big[\hat W_0,{\p\over\p p_1}\Big],\ \ \ \ \
\hat F_2=\Big[\hat W_0,\hat F_1\Big]
\ee
With these operators, one constructs recurrently
\be
\hat W_{m+1}={1\over m}\Big[\hat W_m,\hat F_2\Big]
\ee
with the initial condition $\hat W_1=F_1$.
Thus, finally, the partition function (\ref{Z1}), or, equivalently, the matrix model (\ref{im}) is generated by these operators
\be
Z(N;\bar p,p,g)=\exp\left(\sum_{k=1}{\bar p_m\hat W_m[p]\over m}\right)\cdot \exp\left(\sum_{k=1}{p_kg_k\over k}\right)
\ee

Similarly, operators $\hat W_{-m}[p]$ from (\ref{W-m}) are also constructed using three auxiliary operators: the same cut-and-join operator (\ref{W0}) and operators
\be
\hat E_1=\Big[\hat W_0,p_1\Big],\ \ \ \ \
\hat E_2=\Big[\hat W_0,\hat E_1\Big]
\ee
With these operators, one constructs recurrently
\be
\hat W_{-m}={1\over m-1}\Big[\hat E_2, \hat W_{-m+1}\Big]
\ee
with the initial condition $\hat W_{-1}=E_1$.
Thus, finally, the partition function (\ref{Z1}), or, equivalently, the matrix model (\ref{nb}) is generated by these operators
\be
Z_-(N;\bar p,g)=Z(N;\bar p,p=0,g)=\exp\left(\sum_{k=1}{\bar p_m\hat W_{-m}[g]\over m}\right)\cdot 1
\ee

\section{Generalizing the WLZZ series}

Now we are ready to present a simple generalization of the interpolating model discussed in the previous section. This generalization is related with higher roots of the $w_\infty$-algebra. The partition function of this model is specified by choosing a polynomial $f(x)=\prod_{l=1}^n(N_l+x)$:
\be\label{Zn}
Z^{(n)}(N_l;\bar p,p,g)=\sum_{R,Q}\left(\prod_{l=1}^{n}{\prod_{i,j\in R}(N_l+j-i)\over \prod_{i,j\in Q}(N_l+j-i)}\right)S_{R/Q}\{\bar p\}S_R\{g\}S_Q\{p\}
\ee
The WLZZ-models correspond to $n=1$. This partition function celebrates the same integrability properties, however, its matrix model representation is more involved: it depends on a set of integers $N_i$, which are sizes of matrices in the multi-matrix model, which is quite involved even in the case of $p_k=0$, i.e. that corresponding to the negative branch of the WLZZ models: see \cite[Eq.(59)]{AMMN}.

\subsection{$W$-representation}

Though the matrix model representation becomes very involved for the class of models $Z^{(n)}$ as compared with $Z^{(n)}$ in sec.3, their $W$-representations are still of the same complexity. Let us construct the operators $\hat W_{m}^{(n)}[p]$ from (\ref{Wm}). They can be produced from operators already constructed in the previous section.

We will need a whole set of new auxiliary operators $\hat F_k(N_l)$ in addition to the cut-and-join operator (\ref{W0}). These operators depends on the set of integers $N_l$, $l=1,\ldots,k-1$, and are constructed iteratively:
\be
\hat F_{k+1}(N_l)=\Big[\hat W_0(N_l),\hat F_k(N_l)\Big]
\ee
With these operators, one constructs recurrently the set
\be
\hat W_{m+1}^{(n)}={1\over m}\Big[\hat W_{m}^{(n)},\hat F_{n+1}\Big]
\ee
with the initial condition $\hat W_{1}^{(n)}=F_n$.

Thus, finally, the partition function (\ref{Z1}), or, equivalently, the matrix model (\ref{im}) is generated by these operators
\be
Z^{(n)}(N_l;\bar p,p,g)=\exp\left(\sum_{k=1}{\bar p_m\hat W_m^{(n)}[p]\over m}\right)\cdot \exp\left(\sum_{k=1}{p_kg_k\over k}\right)
\ee

Similarly, operators $\hat W_{-m}^{(n)}[p]$ from (\ref{W-m}) are also constructed using a set of auxiliary operators: this time they are
\be
\hat E_{k+1}(N_l)=\Big[\hat W_0(N_1),\hat E_k(N_l)\Big]
\ee
With these operators, one again constructs recurrently the set
\be
\hat W_{-m}^{(n)}={1\over m-1}\Big[\hat E_{n+1}, \hat W_{-m+1}^{(n)}\Big]
\ee
with the initial condition $\hat W_{-1}^{(n)}=E_n$.

Thus, finally, the partition function (\ref{Z1}), or, equivalently, the matrix model (\ref{nb}) is generated by these operators
\be
Z_-^{(n)}(N_l;\bar p,g)=Z^{(n)}(N_l;\bar p,p=0,g)=\exp\left(\sum_{k=1}{\bar p_m\hat W_{-m}^{(n)}[g]\over m}\right)\cdot 1
\ee

\subsection{$w_\infty$-algebra}

To bring a little more order in the zoo of considered operators, let us separately discuss their relation to each other. For the sake of simplicity, we put $N_l=0$ in all operators. The motivation for this is that turning on $N$ amounts to the shift
\begin{equation}
   \hat W_0(N) =\hat W_0(0) +  N \, \hat L_0.
\end{equation}
which then translates to other formulas. This does not respect the grading.

At $N=0$, we will drop the dependence on $N$  in the notation.  Then the operators, constructed above can be represented as the graded generators of the $w_\infty$-algebra.

The $w_{\infty}$ algebra is formed by the operators of the form $p_{a_1}\ldots p_{a_{m}}
\frac{\p^n}{\p p_{b_1}\ldots \p p_{b_n}}$, and they
are basically classified by two gradings: the spin, which is
the net number of $p$'s, $n+m$, and the second grading: the sum of indices $b_1+\ldots+b_m-a_1-\ldots-a_n$,
which we plot at the vertical and the horizontal axes respectively.

Let us denote as $V_{(m,n)}$ an operator of spin $m$ and of the second grading $n$. Then:
\be
\hat E_k=V_{(k+1,1)},\ \ \ \ \ p_k=V_{(1,k)},\ \ \ \ \ \hat W_0={1\over 2}V_{(3,0)},\ \ \ \ \hat W_{-m}^{(n)}= V_{(mn+1,m)}
\ee
These operators satisfy the $w_{1+\infty}$ algebra relations \cite{Bakas}:
\be\label{cr}
[V_{(m_1,n_1)},V_{(m_2,n_2)}]=((m_1-1)n_2-(m_2-1)n_1)
V_{(m_1+m_2-2,n_1+n_2)}
\ee
All the $w_\infty$-operators  can be drawn on the following diagram of generators of the Borel subalgebra of $w_\infty$-algebra:

\vspace{0.7cm}

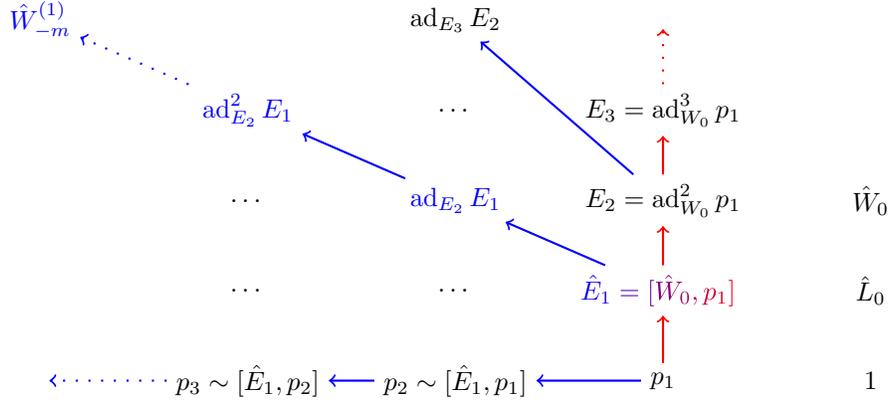
\pgfmathsetmacro{\ra}{2.3}

\begin{center}
\begin{tikzpicture}[scale=1.2]
  \foreach \x [evaluate=\x as \xval using int(\x-1)] in {3,2}
  {
  \node (mx\x) at (-1*\ra*\x,1){ $p_{\x}\sim [\hat E_1,p_{\xval}]$};
  }
  \node (mx1 y0) at (-\ra,1){$p_1$};
	\node (mx4 y0) at (-4*\ra,1) {};
  \node (x0 y1) at (0,1+1) {$\hat L_0$};
  \node (x0 y2) at (0,1+2) {$\hat W_0$};
  \node (mx1 y1) at (-\ra,1+1) {\rbtext{$\hat E_{1}=[\hat W_0,p_1]$}};
  \node (mx2 y1) at (-2*\ra,1+1) {\ldots};
  \node (mx3 y1) at (-3*\ra,1+1) {\ldots};
  \node (mx3 y2) at (-3*\ra,1+2) {\ldots};
  \node (mx2 y3) at (-2*\ra,1+3) {\ldots};
  \node (mx2 y2) at (-2*\ra,1+2) {${\color{blue} \operatorname{ad}_{E_2}E_1 } $ };
\node (mx3 y3) at (-3*\ra,1+3) {$ {\color{blue} \operatorname{ad}^2_{E_2}E_1 }$};
\node (mx1 y2) at (-\ra,1+2) {$E_2=\operatorname{ad}_{W_0}^2 p_1$};
\node (mx1 y3) at (-\ra,1+3) {$E_3=\operatorname{ad}_{W_0}^3 p_1$};
\node (mx1 y4) at (-\ra,1+4) {};
\node (mx2 y4) at (-2*\ra,1+4) {$\operatorname{ad}_{E_3} E_2$};
\node (mx4 y4) at (-4*\ra,1+4) {{\color{blue}$\hat W^{(1)}_{-m}$}};
\node (x0 y0) at (0,1) {$1$};
\draw [blue,thick,->] (mx1 y1) -- (mx2 y2);\draw [blue,thick,->] (mx2 y2) -- (mx3 y3);
\draw [red,thick,->] (mx1 y0) -- (mx1 y1);\draw [red,thick,->] (mx1 y1) -- (mx1 y2);\draw [red,thick,->] (mx1 y2) -- (mx1 y3);
\draw [blue,thick,->] (mx1 y2) to  (mx2 y4);
\draw [blue,thick,->] (mx1 y0) -- (mx2);
\draw [blue,thick,->] (mx2) -- (mx3);
\draw [blue,thick,loosely dotted, <-] (mx4 y4) to (mx3 y3);
\draw[red, thick, loosely dotted, ->] (mx1 y3) to (mx1 y4);
\draw[blue, thick, loosely dotted, ->] (mx3) to (mx4 y0);
\end{tikzpicture}
\end{center}
\vspace{0.7cm}
Moving to the upwards and left is achieved by taking iterated commutators with $\hat E_{n+1}$. Hence, each line of {\color{blue}blue operators} are building blocks of the $Z_{-m}^{(n)}$-series at a concrete $n$. The more we move upwards (along the {\color{red} red arrows}), commuting iteratively with $\hat W_0$, the higher $n$ we choose.

The operators of our interest here occupy not that much part of the table, though one can generate the whole table using the commutation relations (\ref{cr}).

One can construct a similar picture for the positive branch, identifying $k{\p\over\p p_k}=V_{(1,-k)}$, $\hat F_{m}=V_{(1-m,1)}$, etc.
Unfortunately, this version of the $w_\infty$-algebra does not allow a central extension except for the Virasoro generators $V_{(2,k)}$. At the same time, $k{\p\over\p p_k}=V_{(1,-k)}$ does not commute with $V_{(1,k)}=p_k$. Hence, one can not sew the two branches together. This is not surprising, since the corresponding $W$-representations act as differential operators on different sets of variables: $p_k$ and $g_k$. To have a unique picture, one has to construct a $W$-representation of $Z^{(n)}(N_l;\bar p,p,g)$ in terms of differential operators acting on the variables $g_k$. Unfortunately, such a formulation is not known so far.

From the commutation relations (\ref{cr}), one can again check that the infinite families of operators $W_{-m}^{(n)}=V_{(mn+1,m)}$ at each $n$ are commutative, and similarly for  $W_{m}^{(n)}$. Let us introduce an operator $\hat O(N)$ that has the Schur functions as its eigenfunctions:
\be
\hat O(N)\cdot S_R=\left(\prod_{i,j\in R}(N+j-i)\right)\cdot S_R
\ee
Such an operator has been manifestly constructed in \cite[Eqs.(21),(26)]{AMMN}. As is clear from (\ref{W-m}) and (\ref{scp}), the operators $W_{-m}^{(1)}$ can be constructed using this operator: 
\be
W_{-m}^{(1)}=\hat O(N)\cdot p_m\cdot\hat O^{-1}(N)
\ee
and, more generally, 
\be
W_{-m}^{(n)}=\left(\prod_l^n\hat O(N_l)\right)\cdot p_m\cdot\left(\prod_l^n\hat O(N_l)^{-1}\right)
 \ee
Hence, the powers of the operator $\hat O(N)$ just provide automorphisms of the $w_\infty$-algebra that map the commutative family $\{p_m\}$ to $W_{-m}^{(n)}$, each automorphism producing a blue line in the picture. The vertical line: $1$, $\hat L_0$, $\hat W_0$ is also commutative, and, in terms of the generalized cut-and-join operators $\hat W_\Delta$ of \cite{MMN} is just $\hat W_{[m]}$.

A similar picture is also correct for the $W_{m}^{(n)}$ operators, which are generated from ${\p\over\p p_m}$ with rotation by the same operator $\hat O(N)$ : $W_{m}^{(1)}=\hat O(N)^{-1}\cdot m{\p\over\p p_m}\cdot\hat O(N)$, etc.

\section{Conclusion}

This letter describes an important step in the study of non-perturbative partition functions.
It brings the new class of WLZZ models \cite{WLZZ} into accordance with the traditional
matrix model approach, by suggesting a two-matrix model representation for them.
At the present letter, the evidence is provided just by computer calculation and comparison of averages,
provided by integrals and by the $W$-representations, and by obvious reductions to
simpler one-matrix and two-matrix integrals at distinguished points in the space of parameters, while a
direct derivation will be explained elsewhere.
Now the road is open for application of the standard matrix-model techniques,
which, however, should make new twists because of the broken relation \cite{MMsc} between
the Virasoro/W-based topological recursion \cite{AMM,EO}
and the standard $1/N$ topological expansion \cite{TE}.

We also explained integrability \cite{UFN3} and superintegrability \cite{MMsi} of the new matrix models
by relating them to a new class of $\tau$-functions, which we named skew hypergeometric.
Most important, we now have a unified description of a huge variety of matrix models,
and unification is achieved in terms of the $w_\infty$-algebra, as was long expected.
It is now clear that the crucial lacking point was the need to switch to two-matrix models (this was anticipated many years ago in \cite{Gava,AS}) and, for this, to find an adequate class among them,
which appeared to include the mixture of two ordinary potentials and the Kontsevich background field (in the character phase \cite{GKMU}).
This unifying model with three independent sets of time variables should now be carefully
investigated and extended in various directions.

\section*{Acknowledgements}

We are grateful to D. Galakhov, N. Tselousov and A. Zhabin for stimulating discussions.
Our work is partly supported by the grant of the Foundation for the Advancement of Theoretical Physics ``BASIS" and by the joint grant 21-51-46010-ST-a, and by the National Natural Science Foundation of China (Nos. 11875194 and 12205368).

\begin{appendices}
\section{Explicit formulas for $W$-operators}

In the Appendices, we give explicit expressions for some of the constructed operators,
briefly describe action of various operators on the Schur functions, and also comment on obtaining the Schur function expansion for the partition functions from $W$-operators following the lines of \cite{WLZZ1,MO}.

First of all, we list explicit expressions for first few operators. To begin, we repeat the diagonal operator $W_0(N)$ \eqref{W0}:
\begin{equation}
\hat W_0(N) =\dfrac{1}{2} \sum_{a, b}\left((a+b) p_a p_b \frac{\partial}{\partial p_{a+b}}+a b p_{a+b} \frac{\partial^2}{\partial p_a \partial p_b}\right) + N\sum_k kp_k \dfrac{\partial}{\partial p_k}
\end{equation}
where:
\begin{equation}
	L_0=\sum_k k p_k \dfrac{\partial}{\partial p_k}
\end{equation}
The positive branch is generated by the operators:
\begin{equation}
	\begin{split}
		\hat E_1 &=   \sum_{a} ap_{a+1}\frac{\p}{\p p_{a}} + Np_1  \\
		\hat E_2 &= \sum_{a,b} \left(abp_{a+b+1}\frac{\p^2}{\p p_a\p p_b} + (a+b-1)p_ap_b\frac{\p}{\p p_{a+b-1}}\right)
		+2N\sum_a ap_{a+1}\frac{\p}{\p p_{a}} + N^2p_1
	\end{split}
\end{equation}
where the $N$ dependent terms come from the $L_0$ shifts. This is in agreement with $E_1$ being linear in $N$  as it is a given by a single commutator, while $E_2$ is quadratic as it given by a double commutator. We can clearly see a mix of operators with different spins and the restoration of homogeneity in the spin after formally assigning spin 1 to the parameter $N$. Further operators get increasingly complicated, we list just one representative for an illustration:
\begin{equation}
	\hat{W}_{-2}=[\hat{E}_2,\hat{W}_{-1}]=\sum_{a,b}\left( p_{a+b+2}\dfrac{\partial^2}{\partial p_a \partial p_b}+(a+b-2) p_a p_b \dfrac{\partial}{\partial p_{a+b-2}} \right) + 2N \sum a p_{a+2} \dfrac{\partial}{\partial p_{a}} + N^2 p_2+N p_1^2
\end{equation}
This is the $W$-representation operator for the Gaussian Hermitian matrix model.
\\\\
The operators from the positive branch are given by:
\begin{equation}
\begin{split}
		\hat{F}_1&= - \sum_b (b+1)p_b \dfrac{\partial}{\partial p_{b+1}}-N \dfrac{\partial}{\partial p_1} \\
		\hat{F}_2&=  \sum_{a,b} \left( p_a p_b (a+b+1) \dfrac{\partial }{\partial p_{a+b+1}} + a b p_{a+b-1} \dfrac{\partial^2}{\partial p_a \partial p_b} \right) + 2N \sum (b+1)p_b \dfrac{\partial }{\partial p_{b+1}} +N^2 \dfrac{\partial}{\partial p_1}
\end{split}
\end{equation}

\section{Partition functions from action on the Schur functions}

Action of the operators described in Appendix A on the Schur functions looks as follows. The $\hat{W}_0(N)$-operator acts diagonally:
\begin{equation}
	\hat{W}_0(N) S_R = \left(\sum_{(i,j) \in R} (j-i+N) \right) S_R
\end{equation}
From this, one easily finds:
\begin{equation}
\begin{split}
		p_1 S_R &= \sum_{R+\Box}  S_{R+\Box}\\
		\hat{E}_1 S_R&= \sum_{R+\Box} (j_\Box-i_\Box+N) S_{R+\Box}\\
		\hat{E}_2 S_R&= \sum_{R+\Box} (j_\Box-i_\Box+N)^2 S_{R+\Box}
\end{split}
\end{equation}
where the sum goes over all possible ways to add a single box to the partition $R$. The next series of operators of degree $2$ and of increasing spin would go as:
\begin{equation}
\begin{split}
	p_2 S_R&=[\hat{E}_1,p_1] S_R=\sum_{R+\Box_1+\Box_2} \left\langle p_2 S_R\Big| S_{R+\Box_1+\Box_2}\right\rangle S_{R+\Box_1+\Box_2}
	\\
	\hat{W}_{-2} S_R&=[\hat{E}_2,\hat{E}_1] S_R=\sum_{R+\Box_1+\Box_2} \left\langle p_2 S_R\Big| S_{R+\Box_1+\Box_2}\right\rangle (j_{\Box_1}-i_{\Box_1}+N)(j_{\Box_2}-i_{\Box_2}+N) S_{R+\Box_1+\Box_2}
	\\
	\hat{W}^{(2)}_{-2} S_R&=[\hat{E}_2,\hat{E}_1] S_R=\sum_{R+\Box_1+\Box_2} \left\langle p_2 S_R\Big| S_{R+\Box_1+\Box_2}\right\rangle (j_{\Box_1}-i_{\Box_1}+N)^2(j_{\Box_2}-i_{\Box_2}+N)^2 S_{R+\Box_1+\Box_2}
\end{split}
\end{equation}
The coefficient $\left\langle p_2 S_R\Big| S_{R+\Box_1+\Box_2}\right\rangle$ is in fact quite simple, and is given by:
\begin{equation}
	\left\langle p_2 S_R\Big| S_{R+\Box_1+\Box_2}\right\rangle = \left\{ \begin{split}
		1& \ , \ j_{\Box_2}=j_{\Box_1}, i_{\Box_2}=i_{\Box_1}+1
		\\
		-1& \ , \ j_{\Box_2}=j_{\Box_1}+1, i_{\Box_2}=i_{\Box_1}
	\end{split} \right.
\end{equation}
and vanishes otherwise.
\\\\
Similar formulas hold for the lowering operators:
\begin{equation}
	\begin{split}
		\dfrac{\partial}{\partial p_1} S_R &= \sum_{R-\Box}  S_{R-\Box}\\
		\hat{F}_1 S_R&= \sum_{R+\Box} (j_\Box-i_\Box+N) S_{R-\Box}\\
		\hat{F}_2 S_R&= \sum_{R+\Box} (j_\Box-i_\Box+N)^2  S_{R-\Box}
	\end{split}
\end{equation}
\\\\
Note that the spin grading of a given operator corresponds to the power of $(j-i)$ in terms of its action on the Schur functions.

\section{Obtaining (\ref{Zf}) and (\ref{Zfg}) from the $W$-representations}

Now let us briefly recollect how to obtain the Schur function expansion for partition functions \eqref{Z+f} and \eqref{Z+fg} from \eqref{Wm} and \eqref{W-m} respectively.  This  is just a straightforward calculation. First, consider the negative branch, and expand the exponential in the $W$-representation (\ref{Z+fg}):
\begin{equation}
\begin{split}
	Z_f(\bar p,p=0,g)&=\exp\left(\sum_{k=1}{\bar p_m\hat{W}^f_{-m}[g]\over m}\right)\cdot 1 = \\
	&= \sum_\Delta \dfrac{\bar{p}_\Delta}{z_\Delta} \hat{W}^f_{-\Delta}[g] \cdot 1 = \sum_\Delta \dfrac{\bar{p}_\Delta}{z_\Delta} \sum_R \left(\prod_{(i,j) \in R} f(j-i) \right) \left\langle g_\Delta\cdot 1 \Big| S_R\right\rangle S_R\{g\} =
	\\
	&=  \sum_R \prod_{(i,j) \in R} f(j-i) S_R \{g\} \sum_\Delta \dfrac{\bar{p}_\Delta}{z_\Delta}	\left\langle g_\Delta \cdot 1 \Big| S_R\right\rangle   = \sum_{R}\prod_{i,j\in R} f(j-i)S_{R}\{\bar p\}S_R\{g\}
\end{split}
\end{equation}
where we used formulas (\ref{scp}) and the notation
\begin{equation}
\hat{W}^f_{-\Delta}[g]= \prod_{i=1}^{l(\Delta)} \hat{W}^f_{-\Delta_i}[g] \quad
\end{equation}
For the positive branch, one has:
\begin{equation}
	\begin{split}
		Z_f(\bar p,p,g)&=\exp\left(\sum_{k=1}{\bar p_m\hat W^f_m[p]\over m}\right)\cdot \exp\left(\sum_{k=1}{p_kg_k\over k}\right) =\\
		&=\sum_\Delta \dfrac{\bar{p}_\Delta}{z_\Delta} \hat{W}^f_{\Delta}[g] \cdot \sum_R S_R\{p\} S_R\{g\} =\sum_R S_R \{g\} \sum_{R'} \left(\prod_{(i,j)\in R/Q} f(j-i) \right) S_{Q}(p) \sum_\Delta  \dfrac{\bar{p}_\Delta}{z_\Delta} \left\langle \dfrac{\partial S_R}{\partial p_\Delta} \Big| S_Q \right \rangle =\\
		&= \sum_{R,Q}\prod_{i,j\in R/Q} f(j-i)S_{R/Q}\{\bar p\}S_R\{g\}S_Q\{p\}
	\end{split}
\end{equation}
where we again used formulas (\ref{scp}) and $ \dfrac{\partial}{\partial p_\Delta}=\prod_{i=1}^{l(\Delta)} \Delta_i \dfrac{\partial}{\partial p_{\Delta_i}}  \ , \  \hat{W}^f_{\Delta}[p]= \prod_{i=1}^{l(\Delta)} \hat{W}^f_{\Delta_i}[p]$.
\end{appendices}


\begin{thebibliography}{12}

\bibitem{WLZZ} R.~Wang, F.~Liu, C.~H.~Zhang and W.~Z.~Zhao,
Eur. Phys. J. C \textbf{82} (2022) 902,
arXiv:2206.13038

\bibitem{MMsi} A.~Mironov and A.~Morozov,
Phys. Lett. B \textbf{835} (2022) 137573,
arXiv:2201.12917

\bibitem{UFN3} A. Morozov,
Phys.Usp.(UFN) {\bf 37} (1994) 1;
hep-th/9502091; hep-th/0502010\\
A. Mironov, Int.J.Mod.Phys. {\bf A9} (1994) 4355; Phys.Part.Nucl.
{\bf 33} (2002) 537; hep-th/9409190

\bibitem{MMsc} A.~Mironov and A.~Morozov,
arXiv:2210.09993

\bibitem{AMM} A. Alexandrov, A. Mironov, A. Morozov,
Physica {\bf D235} (2007) 126-167, hep-th/0608228; 
Theor. Math. Phys. \textbf{150} (2007) 153-164,
hep-th/0605171;
JHEP {\bf 12} (2009) 053, arXiv:0906.3305

\bibitem{EO} L. Chekhov and B. Eynard,
JHEP \textbf{0603} (2006) 014, hep-th/0504116;
JHEP \textbf{0612} (2006) 026, math-ph/0604014; \\
 B. Eynard, N. Orantin, Commun. Number Theory Phys. {\bf 1} (2007) 347-452, math-ph/0702045\\
N. Orantin,
arXiv:0808.0635

\bibitem{AMMdop} A. Alexandrov, A. Mironov and A. Morozov,
Int.J.Mod.Phys. {\bf A19} (2004) 4127, hep-th/0310113

\bibitem{UT} K. Ueno, K.Takasaki  Adv.Studies in Pure Math. {\bf 4} (1984) 1

\bibitem{GKM2} S. Kharchev, A. Marshakov, A. Mironov, A. Morozov, Int.J.Mod.Phys. {\bf A10} (1995) 2015, hep-th/9312210

\bibitem{OS} A. Orlov and D.M. Shcherbin,
Theor.Math.Phys.
{\bf 128} (2001) 906-926

\bibitem{Mac} I.G. Macdonald,
{\it Symmetric functions and Hall polynomials}, Second Edition, Oxford University Press,
1995

\bibitem{AMMNi} A.~Alexandrov, A.~Mironov, A.~Morozov and S.~Natanzon,
J. Phys. A \textbf{45} (2012) 045209,
arXiv:1103.4100

\bibitem{Taka} K.Takasaki, Adv.Studies in Pure Math. {\bf 4} (1984) 139-163

\bibitem{DJKM} E.Date, M.Jimbo, M.Kashiwara and T.Miwa, {\sl Transformation
groups for soliton equations}, RIMS Symp. {\sl "Non-linear integrable
systems -- classical theory and quantum theory"} (World Scientific,
Singapore, 1983)

\bibitem{wrep} A. Givental, 
math.AG/0008067

\bibitem{wrep1} A. Alexandrov, A. Mironov, A. Morozov,
Physica {\bf D235} (2007) 126-167, hep-th/0608228\\
A.~~Alexandrov, A.~Mironov, A.~Morozov,
Theor. Math. Phys. \textbf{150} (2007) 153-164,
hep-th/0605171

\bibitem{wrep2} A.Okounkov,
Math.Res.Lett. {\bf 7}
(2000) 447-453;\\
V.Bouchard, M.Marino,
In: {\sl From Hodge Theory to Integrability and tQFT: tt*-geometry},
Proceedings of Symposia in Pure Mathematics, AMS (2008), arXiv:0709.1458;\\
S.Lando,
In: {\sl Applications of Group Theory to Combinatorics}, Koolen,
Kwak and Xu, Eds.
Taylor \& Francis Group, London, 2008, 109-132;\\
M.Kazarian,
arXiv:0809.3263;\\
A.Mironov, A.Morozov,
JHEP \textbf{0902} (2009) 024, arXiv:0807.2843

\bibitem{MSh} A.~Morozov, S.~Shakirov,
  JHEP {\bf 0904} (2009) 064,
arXiv:0902.2627

\bibitem{Wmore} A. Alexandrov, Mod.Phys.Lett. {\bf A26} (2011) 2193-2199, arXiv:1009.4887\\
A.~Alexandrov,
  Adv.Theor.Math.Phys.\  {\bf 22} (2018) 1347,
arXiv:1608.01627\\
H.~Itoyama, A.~Mironov, A.~Morozov,
  JHEP {\bf 1706} (2017) 115,
arXiv:1704.08648

\bibitem{Max} L.~Cassia, R.~Lodin and M.~Zabzine,
Commun. Math. Phys. \textbf{387} (2021) 1729-1755,
arXiv:2102.05682

\bibitem{MMM} A.~Mironov, V.~Mishnyakov, A.~Morozov and R.~Rashkov,
Eur. Phys. J. C \textbf{81} (2021) 1140,
arXiv:2105.09920\\
A.~Mironov, V.~Mishnyakov and A.~Morozov,
Phys. Lett. B \textbf{823} (2021) 136721,
arXiv:2107.02210

\bibitem{AMMN} A.~Alexandrov, A.~Mironov, A.~Morozov and S.~Natanzon,
JHEP \textbf{11} (2014) 080,
arXiv:1405.1395

\bibitem{Orlov} A.Orlov, Theor.Math.Phys. {\bf 146} (2006) 183-206

\bibitem{Al} A.~Alexandrov,
arXiv:2212.10952

\bibitem{WLZZ1} R.~Wang, C.~H.~Zhang, F.~H.~Zhang and W.~Z.~Zhao,
Nucl. Phys. B \textbf{985} (2022) 115989,
arXiv:2203.14578

\bibitem{GJ} D. Goulden, D.M. Jackson, A. Vainshtein,
Ann. of Comb. {\bf 4} (2000) 27-46,
Brikh\"auser, math/9902125

\bibitem{MMN} A.~Mironov, A.~Morozov and S.~Natanzon,
JHEP \textbf{11} (2011) 097,
arXiv:1108.0885;
J. Geom. Phys. \textbf{62} (2012) 148-155,
arXiv:1012.0433

\bibitem{Bakas} I. Bakas, Phys. Lett. {\bf B228} (1989) 57-63\\
H.~Awata, M.~Fukuma, Y.~Matsuo and S.~Odake,
Prog. Theor. Phys. Suppl. \textbf{118} (1995) 343-374,
hep-th/9408158

\bibitem{TE} D. Bessis,
Commun.Math.Phys. {\bf 69} (1979) 147\\
D. Bessis, C. Itzykson and J.B. Zuber,
Adv.Appl.Math. {\bf 1} (1980) 109\\
C. Itzykson and J.B. Zuber,
J.~Math.Phys. {\bf 21} (1980) 411

\bibitem{Gava} A.~Marshakov, A.~Mironov and A.~Morozov,
Mod. Phys. Lett. A \textbf{7} (1992) 1345-1360,
hep-th/9201010

\bibitem{AS} C.R.~Ahn and K.~Shigemoto,
Phys. Lett. B \textbf{285} (1992) 42-48,
hep-th/9112057

\bibitem{GKMU} A.~Mironov, A.~Morozov, G.~W.~Semenoff,
  Int.\ J.\ Mod.\ Phys.\ {\bf A11} (1996) 5031,
  hep-th/9404005

\bibitem{MO} V.~Mishnyakov and A.~Oreshina,
Eur. Phys. J. C \textbf{82} (2022) 548,
arXiv:2203.15675

\end{thebibliography}
\end{document}